\documentclass{ws-ijmpa}

\begin{document}

\markboth{M.~D'ELIA, A.~DI~GIACOMO, C.~PICA}
{ON THE ORDER OF THE DECONFINING TRANSITION IN $N_f=2$ QCD}

%
\catchline{}{}{}{}{}
%

\title{ON THE ORDER OF THE DECONFINING\\ TRANSITION IN $N_f=2$ QCD\footnote{Presented by A.~Di~Giacomo at the $8^{th}$ workshop on non perturbative QCD - Paris, June 7-11, 2004.}}

\author{\footnotesize M.~D'ELIA}
\address{Dipartimento di Fisica, Universit\`a di Genova and INFN\\
via Dodecanneso 33, I-16146, Genova, Italy}

\author{\footnotesize A.~DI~GIACOMO}
\author{\footnotesize C.~PICA}
\address{Dipartimento di Fisica, Universit\`a di Pisa and INFN\\
l.go Pontecorvo 2, I-56127, Pisa, Italy}

\maketitle


\begin{abstract}
A careful study is made on the lattice of the phase diagram of QCD with two staggered flavors, to investigate the order of the chiral transition of $N_f=2$ QCD.
The specific heat and the susceptibility of the chiral condensate are
determined for different spatial sizes of the system, and a finite size scaling
analysis provides a determination of the (pseudo)critical indices.
The result  is a strong indication that the chiral transition is first order.

\keywords{Confinement; Lattice QCD; Chiral transition.}
\end{abstract}

\section{Introduction - Motivation}

\begin{figure}[bt]
\center\includegraphics*[height=3cm]{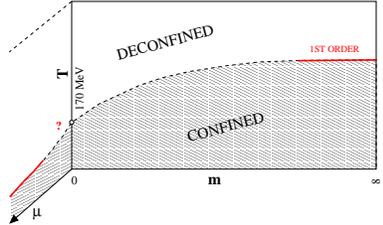}
\caption{Schematic phase diagram of $N_f=2$ QCD.}\label{PHDIA}
\end{figure}

QCD with $N_f=2$ is a specially interesting system to understand confinement.
A schematic view of the phase diagram is shown in Fig.~\ref{PHDIA}, where for simplicity the two quark masses have been put equal to $m$ and $\mu$ is the barion chemical potential..
At $\mu=0$ the transition line between the region which is conventionally named ``confined'' and the ``deconfined'' one is defined by the maximum of a number of susceptibilities ($C_V$, $\chi_{\bar\psi\psi}$, \dots) which happen to coincide within errors.
As $m\rightarrow\infty$ the quenched case is recovered, and the transition is known to be first order, with $\langle L\rangle$, the Polyakov loop, as order parameter, and $Z_3$ as symmetry.
In principle $Z_3$ is explicitly broken by the coupling to the quarks and $\langle L\rangle$ is not an order parameter: however the quenched description is valid empirically down to $m\simeq 2.5-3$ $GeV$.
At $m\simeq 0$ a chiral phase transition takes place, from the low temperature phase where chiral symmetry is spontaneously broken to a phase in which it is restored: there $\langle\bar\psi\psi\rangle$ is a good order parameter.
The mass term explicitly breaks chiral symmetry but $\langle\bar\psi\psi\rangle$ is expected to work as an order parameter in a neighborhood of $m=0$ .

At intermediate values of $m$ neither $\langle L\rangle$ nor $\langle\bar\psi\psi\rangle$ are expected to be good order parameters, even if their susceptibilities show a maximum along the transition line of Fig.~\ref{PHDIA}.
At $m\simeq 0$ another phase transition is expected to take place, with restoration of the $U_A(1)$ symmetry, which is broken by the anomaly at low temperatures.
An independent definition of confinement is needed to ask the question whether deconfinement transition coincides with chiral and $U_A(1)$ transition.
An effective critical free energy density $L_\phi$ can be written for the chiral transition\cite{piwi}, assuming that the relevant critical excitation are the scalar and pseudoscalar particles.
\begin{equation}
\tilde\phi: \ \ \phi_{ij} = \langle \bar q_i (1+\gamma_5) q_j\rangle \ \ \ \ (i,j = 1..N_f)
\end{equation}
Under chiral and $U_A(1)$ transformations of the group $U_A(1)\otimes SU(N_f)\otimes SU(N_f)$, $\tilde\phi$ transforms as
\begin{equation}
\tilde\phi \rightarrow e^{i\alpha} U_+ \tilde\phi U_-
\end{equation}
so that by the usual symmetry arguments, and neglecting irrelevant terms
\begin{equation}
L_\phi = \frac{1}{2} Tr \left[ \partial_\mu\phi^\dagger\partial^\mu\phi \right] - \frac{m_\phi^2}{2} Tr \{ \phi^\dagger\phi\} - \frac{\pi^2}{3} g_1 \left[ Tr \{ \phi^\dagger\phi \}\right]^2 - \frac{\pi^2}{3} g_2 Tr \{ \phi^\dagger\phi\}^2.
\end{equation}
Inclusion of anomaly brings in an additional term $L_\phi ' = c\left[ det\phi + det\phi^\dagger\right]$, which is $SU(N_f)\otimes SU(N_f)$ invariant, but not $U_A(1)$ invariant.

One can inquire, by use of renormalization group plus $\epsilon$-expansion techniques, if infrared stable fixed points exist, which indicate the possible existence of a second or higher order transition.
For $N_f\geq 3$ no such point exists, and the chiral transition is first order.
At $N_f=2$, in the absence of anomaly ($c=0$) or if the $\eta '$ mass vanishes at $T_c$, no fixed point exists, the transition is first order, and also at $m$, $\mu\neq 0$ the transition is expected to be first order.
In such a case no tricritical point exists in the plane $(\mu,T)$ of Fig.~\ref{PHDIA}\cite{TRICRI}.
If instead $m_{\eta '}\neq 0$ at $T_c$, and the $U_A(1)$ transition occurs at $T>T_c$, the symmetry group is $O(4)$ and the transition can be second order.
In that case at $m$, $\mu\neq 0$ there is no transition but only a crossover, and a tricritical point is expected in the $(\mu,T)$ plane of Fig.~\ref{PHDIA}.

The issue has fundamental implications for confinement.
If the deconfining transition is order-disorder, so that an order parameter exists, a crossover is excluded.
If instead a crossover exists, a state of free quarks can continously be deformed to the confined phase, and can exist also there.
The existing literature\cite{colombia,karsch,jlqcd,bernard} is not conclusive on this point, even if a preference is given to the second order plus crossover scenario.
The problem deserves more attention.
We are going to present preliminary results based on a 5~$TFlops\times month$ lattice simulations on $APEmille$ computer.

\section{Lattice Investigation}

The order of the chiral transition, and, more generally, the transition line of Fig.~\ref{PHDIA}, can be studied by lattice simulations and standard finite size scaling techniques.
We have used staggered fermions on $L_t\times L_s^3$ lattices, with $L_t=4$ and $L_s=12,16,20,24,32$.
The input parameters of the simulations are $\beta=6/g^2$ and $am$, the quark mass in units of the inverse lattice spacing $a^{-1}$.
The temperature is given by
\begin{equation}
T = \frac{1}{L_t a(\beta, m)}
\end{equation}
with $a(\beta, m)$ the lattice spacing in physical units.
The reduced temperature $\tau\equiv ( 1- T/T_c)$ is then given by
\begin{equation}
\tau = 1- \frac{a(\beta_c, 0)}{a(\beta, m)}
\end{equation}
or, in a sufficiently small neighborhood of the critical point
\begin{equation}
\tau = \frac{\partial\ln a}{\partial\beta}\Big|_{(\beta_c,0)} \left[\beta_c - \beta + k m\right]
\end{equation}
with
\begin{equation}
k = \frac{\frac{\partial\ln a}{\partial m}}{\frac{\partial\ln a}{\partial\beta}}\Big|_{(\beta_c,0)}
\end{equation}
In the quenched case $k=0$ and $\tau \propto (\beta_c - \beta)$.

The correlation lenght of the order parameter $\xi$ diverges at the critical point
\begin{equation}
\xi \rightarrow \tau^{-\nu}\ \ for\ \ \tau\rightarrow 0^+
\end{equation}
with a critical index known as $\nu$.
For the specific heat $C_V$ and for the susceptibility $\chi$ of the order parameter, the following scaling laws are expected
\begin{eqnarray}
C_V-C_0 &=& L_s^{\alpha/\nu}\Phi_C(\tau L_s^{1/\nu}, am_q L_s^{y_h} )\label{CV1} \\
\chi -\chi_0 &=& L_s^{\gamma/\nu}\Phi_\chi(\tau L_s^{1/\nu}, am_q L_s^{y_h} )\label{CHI1}
\end{eqnarray}
Eqs.~\ref{CV1}-\ref{CHI1} are obtained by renormalization group arguments, holding when the lattice spacing is much smaller than $\xi$, so that $a/\xi \approx 0$.
This is true for second order and weak first order transitions.
The critical indices $\nu$, $\alpha$, $\gamma$, $y_h$ depend on the transition (see Table~\ref{CRITEXP}).
\begin{table}[b!]
\tbl{Critical exponents.}{
\begin{tabular}{|c|c|c|c|c|c|}
\hline & $y_t$ & $y_h$ & $\nu$ & $\alpha$ & $\gamma$\\
\hline $O(4)$ & 1.34 & 2.49 & 0.75 & -0.23 & 1.48\\
\hline $O(2)$ & 1.49 & 2.49 & 0.67 & -0.01 & 1.33\\
\hline $MF$ & $3/2$ & $9/4$ & $2/3$ & 0 & 1\\
\hline $1^{st} Order$ & 3 & 3 & $1/3$ & 1 & 1\\
\hline
\end{tabular}}
\label{CRITEXP}
\end{table}
The physics behind all that is that around $T_c$ the free energy can be written in terms of the order parameter, and the dependence is dictated by symmetry: irrelevant (higher dimensional) terms are neglected.
Eq.\ref{CV1} for the specific heat is always valid. Eq.\ref{CHI1} is only valid if the choice of the order parameter is correct.
Therefore the specific heat is a reference to test the validity of any order parameter.

\begin{figure*}[t!]
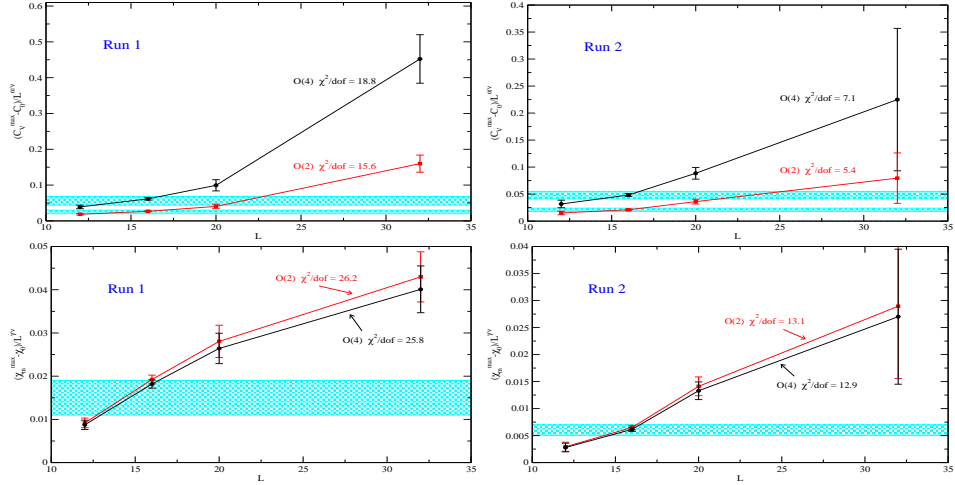

\includegraphics*[width=0.49\textwidth, height=0.25\textwidth]{Cv_max_Run1.eps}
\includegraphics*[width=0.49\textwidth, height=0.25\textwidth]{Cv_max_Run2.eps}\\
\includegraphics*[width=0.49\textwidth, height=0.25\textwidth]{Chi_max_Run1.eps}
\includegraphics*[width=0.49\textwidth, height=0.25\textwidth]{Chi_max_Run2.eps}
\caption{Specific heat (top) and $\chi_m$ (bottom) peak value for Run1 (left) and for Run2 (right), divided by the appropriate powers of $L_s$ (Eqs.~\ref{CV}-\ref{CHI}) to give a constant. Horizontal stripes indicate the $1\sigma$ confidence region for a fit with a constant value. $\chi^2/dof$ is also shown.}\label{R12}
\end{figure*}

In order to test the $O(4)$ option and second order, we have run at fixed value of the scaling variable $am_qL_s^{y_h}$ with $y_h=2.49$, and different spatial sizes $L_s$.
The expectation is then, from Eqs.~\ref{CV1}-\ref{CHI1}, that the peak values scale as
\begin{eqnarray}
(C_V-C_0)^{peak}\propto L_s^{\alpha/\nu} \label{CV}\\
(\chi-\chi_0)^{peak}\propto L_s^{\gamma/\nu}. \label{CHI}
\end{eqnarray}
Since the value of $\alpha$ for $O(4)$ is negative the height of the peak for $C_V-C_0$ is expected to decrease at high $L_s$.
Data show instead a rapid raise.
The quantities $(C_V-C_0)^{peak}/ L_s^{\alpha/\nu}$ and $(\chi-\chi_0)^{peak}/L_s^{\gamma/\nu}$, which should be constant if $O(4)$ were the symmetry, are displayed in Fig.~\ref{R12}.
The $\chi^2/dof$ for a constant is very high: $O(4)$ symmetry is excluded.
Since $y_h$ for $O(2)$ is equal within errors to that of $O(4)$, also $O(2)$ symmetry can be tested and it turns out to be equally bad.
Our action is not ``improved'': however we do not expect that an infrared property like Eqs.~\ref{CV}-\ref{CHI} are affected by ultraviolet improvement.
We can safely state that $O(4)$ and $O(2)$ are excluded and with them the crossover scenario.

Continuity arguments and Eqs.~\ref{CV1}-\ref{CHI1} require that, at small values of $m$
\begin{eqnarray}
C_V-C_0 &\simeq& (am_q)^{-\alpha/(\nu y_h)}\Phi_C(\tau L_s^{1/\nu}, am_q L_s^{y_h} )\label{CV2} \\
\chi -\chi_0 &\simeq& (am_q)^{-\gamma/(\nu y_h)}\Phi_\chi(\tau L_s^{1/\nu}, am_q L_s^{y_h} )\label{CHI2}
\end{eqnarray}
The positions of the peaks $\beta_{max}$ scale then as
\begin{equation}
\beta_c - \beta_{max} + k m - k' L_s^{-1/\nu} = 0 \label{BETAMAX}
\end{equation}
their heights as
\begin{eqnarray}
(C_V-C_0)^{peak} (am_q)^{\alpha/(\nu y_h)} = const \label{CV3} \\
(\chi -\chi_0)^{peak} (am_q)^{\gamma/(\nu y_h)} = const \label{CHI3}
\end{eqnarray}

\begin{figure*}[hbt!]
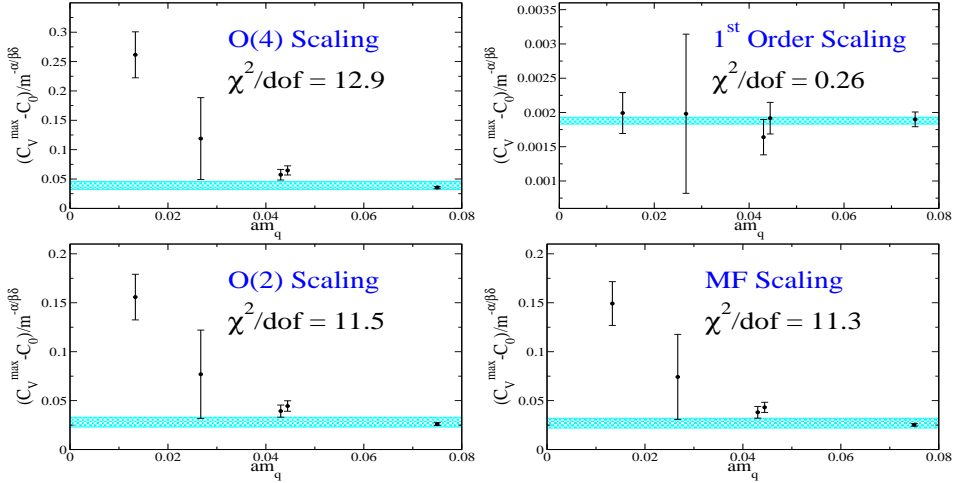

\includegraphics*[width=0.49\textwidth, height=0.25\textwidth]{Cv_max_O4.eps}
\includegraphics*[width=0.49\textwidth, height=0.25\textwidth]{Cv_max_1st.eps}\\
\includegraphics*[width=0.49\textwidth, height=0.25\textwidth]{Cv_max_O2.eps}
\includegraphics*[width=0.49\textwidth, height=0.25\textwidth]{Cv_max_mf.eps}
\caption{Peak of the specific heat divided by the appropriate power of the mass (Eq.~\ref{CV3}) to give a constant, for different scaling hypotheses.}\label{SH}
\end{figure*}

Eqs.~\ref{CV2},~\ref{CHI2},~\ref{BETAMAX},~\ref{CV3}~and~\ref{CHI3} can be tested with the data.
Eq.~\ref{BETAMAX} is compatible both with first order ($\nu=1/3$) and $O(4)$, in agreement with previous analyses\cite{karsch,jlqcd}.
Eqs.~\ref{CV3}-\ref{CHI3} again exclude $O(4)$ and $O(2)$ and are consistent with a first order transition (see Fig~\ref{SH}).

\section{Dual Superconductivity of the Vacuum}
\begin{figure*}[hbt!]
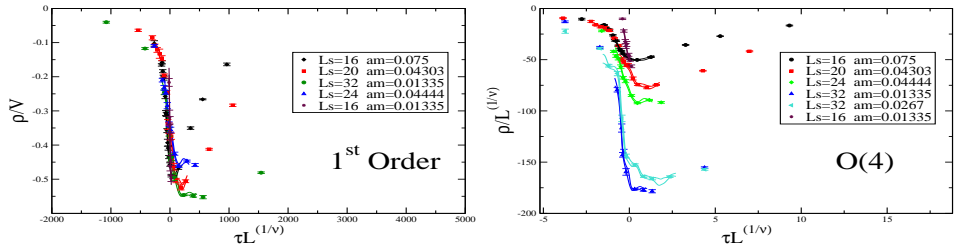

\includegraphics*[width=0.49\textwidth, height=0.25\textwidth]{Rho_su_vol.eps}
\includegraphics*[width=0.49\textwidth, height=0.25\textwidth]{Rho_O4.eps}
\caption{$\rho/L_s^{1/\nu}$ scaling according to Eq.~\ref{RHOSCA} for different scaling hypotheses.}\label{RHOFIG}
\end{figure*}

A disorder parameter $\langle\mu\rangle$ detecting dual superconductivity of the vacuum has been developed by our group during the last years and proved to be a good parameter for the quenched theory.
The parameter can be defined equally well in the presence of dynamical quarks.
$\langle\mu\rangle$ is strictly zero in the deconfine phase of Fig.~\ref{PHDIA} and it drops to zero at the transition line, as shown by the fact that the susceptibility
\begin{equation}
\rho = \frac{\partial}{\partial\beta}\ln\langle\mu\rangle
\end{equation}
has a sharp negative peak coincident within errors with the peak of $C_v$.
Around $T_c$
\begin{equation}
\langle\mu\rangle = L_s^{k/\nu}\Phi_{\langle\mu\rangle}(\tau L_s^{1/\nu}, am_q L_s^{y_h} )
\end{equation}
continuity arguments require
\begin{equation}
\langle\mu\rangle \simeq (am_q)^{-k/(\nu y_h)}\tilde\Phi_{\langle\mu\rangle}(\tau L_s^{1/\nu})
\end{equation}
and therefore
\begin{equation}
\rho/L_s^{1/\nu} \simeq f_{\langle\mu\rangle}(\tau L_s^{1/\nu})\label{RHOSCA}
\end{equation}
Independence on $m$ is expected.
Eq.~\ref{RHOSCA} allows a determination of $\nu$.
If this agrees with the value obtained from $C_v$, a legitimation of $\langle\mu\rangle$ as order parameter follows.
Eq.~\ref{RHOSCA} is compatible with first order (Fig.~\ref{RHOFIG}).

\section{Conclusions}

Finite size scaling analysis of $N_f=2$ QCD definetely excludes $O(4)$ --and $O(2)$-- symmetry at the chiral critical point, and favours a first order transition.
The investigation is continuing with larger lattices and improved actions, to better test the first order option.
Dual superconductivity of the vacuum is confirmed as a good symmetry for the order parameter.

\end{document}